\documentclass[a4paper,11pt]{article}
\usepackage{pos}
\usepackage{graphicx}
\def\logo{} 

\title{The Light Dark Matter eXperiment}

\ShortTitle{The Light Dark Matter eXperiment}

\author*[a]{Tamas Almos Vami}

\affiliation[a]{On behalf of the LDMX Collaboration\\
  University of California, Santa Barbara, Santa Barbara, CA 93106, USA}

\emailAdd{tamas.almos.vami@cern.ch}

\abstract{Dark matter (DM) with a mass between 1\,MeV and 1\,GeV is strongly motivated by thermal
freeze-out scenarios, yet remains largely unexplored by accelerator experiments. The proposed Light Dark Matter
eXperiment (LDMX) will send the 8\,GeV LCLS-II electron beam at SLAC onto a thin target to produce
sub-GeV DM directly, and will identify it by the momentum and energy that the recoiling electron is
missing. These proceedings describe the detector design and the strategy to suppress
standard model backgrounds by 14 orders of magnitude. An updated background rejection study is presented, with
the latest detector geometry, more realistic tracking, and three times more simulated events. A boosted
decision tree using electromagnetic calorimeter observables, combined with a hadronic calorimeter veto,
leaves 1 background event at $1.5\times10^{14}$ electrons on target while retaining 72--79\% of the
signal. A re-optimized graph neural network and cut-and-count
analysis are also compared, and the expected reach
and the ability to characterize a signal are discussed. Finally, the sensitivity to long-lived dark
photons and axion-like particles, and the first beam data recorded with a vertical slice of the
detector, are summarized.}

\FullConference{43rd International Conference on High Energy Physics (ICHEP 2026)\\
30 July  to 5 August , 2026\\
Natal, Brazil\\}

\begin{document}
\maketitle

\section{Light dark matter production}
\label{sec:intro}

Dark matter (DM) searches at the Large Hadron Collider and elsewhere have mostly targeted masses
from 1 GeV to 10 TeV~\cite{cms-summary,atlas-summary,NA64:2019imj}. Between 1\,MeV and 1\,GeV the
parameter space is much less explored, although thermal scenarios motivate it well: DM stays in
equilibrium with the standard model (SM) particles in the early universe until freeze-out sets its
relic abundance~\cite{CatenaGray2023}. The observed abundance then fixes how often a dark photon
($\mathrm{A}'$) is radiated when an electron scatters on a nucleus, in analogy with ordinary
bremsstrahlung. This dark bremsstrahlung is suppressed by roughly 15 orders of magnitude relative to
its SM counterpart, so $10^{14}$--$10^{16}$ electron-target interactions are needed to test thermal DM
conclusively.

At SLAC, the LCLS-II linac delivers 8\,GeV electrons, and End Station A (ESA) is well suited to host a
fixed-target experiment~\cite{markiewicz2022slaclinacesalesa}. The Sector 30 Transfer Line kicker
diverts ultra-low-charge bunches, containing on average one electron, into ESA. There a compact
detector, the Light Dark Matter eXperiment (LDMX), is proposed with a design capable of measuring  every particle in the final state; its
design report is Ref.~\cite{LDMX:2025rwc}.

\section{The LDMX detector}
\label{sec:detector}

\begin{figure}[htbp]
    \centering
    \includegraphics[height=4.5cm]{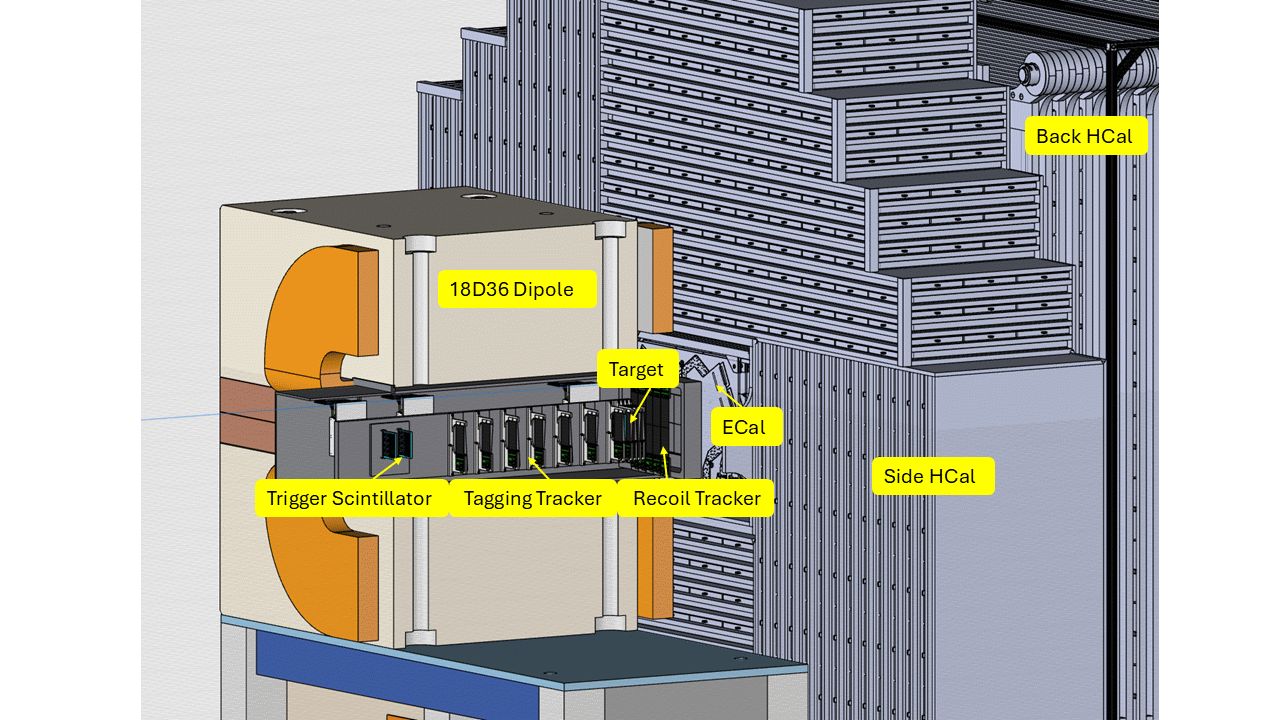}\hspace{0.6cm}%
    \includegraphics[height=4.5cm]{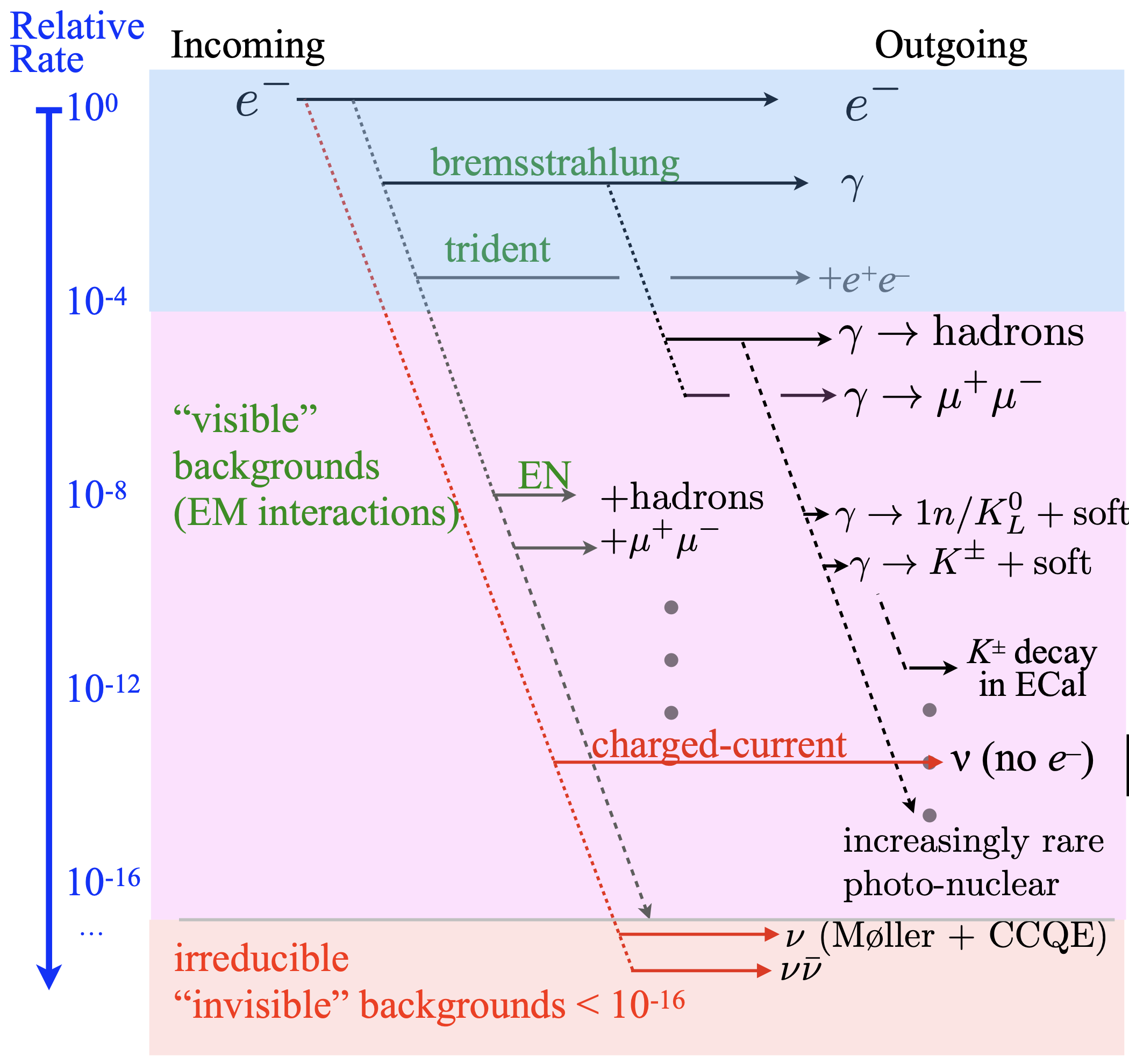}
    \caption{Left: the LDMX detector (trigger scintillators, tagging tracker, target in the dipole,
    recoil tracker, ECal, and HCal). Right: relative rates of the SM backgrounds~\cite{LDMX:2025rwc}.}
    \label{fig:detector}
\end{figure}

A dark bremsstrahlung event leaves a recoil electron with a transverse kick, while the $\mathrm{A}'$
and the DM it decays to escape unseen. LDMX therefore looks for missing momentum and missing energy,
and rejects events in which an SM photon is produced instead. This requires a fast tracker with
precise momentum measurement; a fast, finely segmented, radiation-hard electromagnetic calorimeter; and
a hermetic hadronic calorimeter, each based on technology proven elsewhere
(Fig.~\ref{fig:detector}, left).

The 1.5\,T magnet and the silicon strip trackers are adapted from HPS~\cite{hps_proposal_2014}. Seven
double layers form the tagging tracker, which rejects off-energy beam electrons. They are followed by a
$0.1\,X_0$ tungsten target and by six double layers forming the recoil tracker, which measures the
$p_\mathrm{T}$ of the electron after its transverse kick.

The electromagnetic calorimeter (ECal) scales down the design of the CMS Phase-2 High Granularity
Calorimeter~\cite{Milella:2023tkm,CMSHGCAL:2020dnm}. Silicon sensors are interleaved with 16
tungsten-absorber double layers, and cells of about 5\,mm radius are fine enough to follow minimum
ionizing particles (MIPs) through it. The hadronic calorimeter (HCal) follows the Mu2e cosmic ray
veto~\cite{Mu2e:TDR}: scintillator bars between steel absorbers, read out by SiPMs as in
CMS~\cite{HGCROC}, catch the hadrons, mainly neutrons, that leave the ECal.

\section{Background rejection}
\label{sec:rejection}

At most about one background event in $10^{14}$ may survive, comparable to singling out one cell
among all those in a human body (Fig.~\ref{fig:detector}, right). Bremsstrahlung, the most abundant, is
rejected by the energy of the electron and the photon in the ECal, and photon conversions in the target
by the recoil track multiplicity. The most challenging backgrounds are photonuclear (PN) reactions
producing hadrons in the ECal, which require a detailed understanding of the ECal showers,
the HCal veto, and, for kaons, tracking inside the ECal. M\o ller scattering and processes with
final-state neutrinos matter only beyond the design exposure of LDMX in electrons on target (EoT).

The rejection is established with \textsc{Geant4}~\cite{Agostinelli:2002hh} simulations of
photonuclear and conversion reactions in the ECal and the target (``ECal PN'', ``ECal conv.'',
``target PN'', ``target conv.''), enhanced with a biasing factor that preserves the interaction vertex
position. Dark bremsstrahlung signals with $m_{\mathrm{A}'}$ spanning four orders of magnitude are
simulated with MadGraph~\cite{alwall:2007st} and a dedicated \textsc{Geant4} package~\cite{g4darkbrem}.

\subsection{Updated selections}
\label{sec:bdt}

These proceedings update the results of the design report~\cite{LDMX:2025rwc} and of
Ref.~\cite{Vami:2025phep}. The selections use the latest (v15) detector geometry and more realistic
tracking in the training samples, are evaluated on three times more simulated events, with re-optimized selection criteria. All yields are normalized to
$1.5\times10^{14}$\,EoT. Every selection requires the trigger, i.e.\ less than 3\,GeV reconstructed in
the first 20 ECal layers, and fewer than eight
photoelectrons (PE) in every HCal channel.

\begin{table}[htbp]
\centering
\caption{Cutflow of the BDT selection: background yields at $1.5\times10^{14}$\,EoT and signal efficiency
relative to the acceptance.}
\label{tab:cutflow_bdt}
\resizebox{\textwidth}{!}{%
\begin{tabular}{|l||c|c|c|c||c|c|c|c|} \hline
Selection & \multicolumn{4}{c||}{Number of background events remaining} & \multicolumn{4}{c|}{Signal efficiency for $m_{\mathrm{A}'}$} \\ \hline
 & ECal PN & ECal conv. & target PN & target conv. & 1\,MeV & 10\,MeV & 100\,MeV & 1000\,MeV \\ \hline
All / acceptance            & $3.00\times10^{9}$ & $1.40\times10^{7}$ & $3.96\times10^{5}$ & $1.09\times10^{6}$ & 100\% & 100\% & 100\% & 100\% \\ \hline
Triggered                   & $2.43\times10^{8}$ & $1.07\times10^{7}$ & $9.34\times10^{4}$ & $8.22\times10^{5}$ & 86.4\% & 85.8\% & 86.7\% & 86.5\% \\ \hline
Preselection & 53\,814\,702 & 5\,935\,122 & 57\,498 & 272\,658 & 86.4\% & 85.8\% & 86.7\% & 86.5\% \\ \hline
BDT score $>0.9547$         & 173\,602 & $<$1 & 885 & $<$1 & 74.6\% & 78.7\% & 82.9\% & 84.8\% \\ \hline
HCal max PE $<8$            & 1 & $<$1 & $<$1 & $<$1 & 71.7\% & 75.1\% & 78.7\% & 74.4\% \\ \hline
\end{tabular}}
\end{table}

\begin{figure}[htbp]
    \centering
    \includegraphics[width=0.38\textwidth]{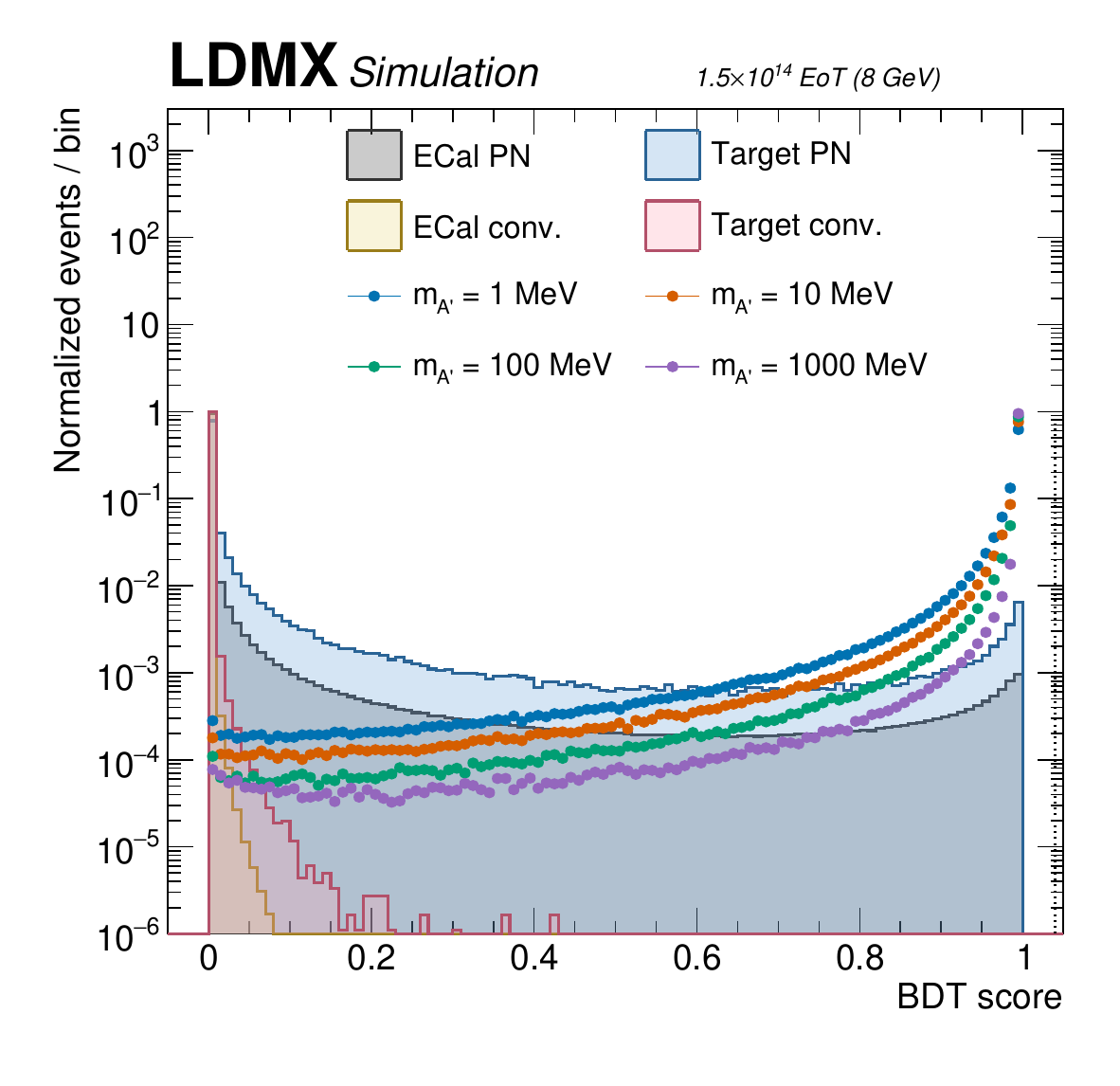}\hspace{0.05\textwidth}
    \includegraphics[width=0.38\textwidth]{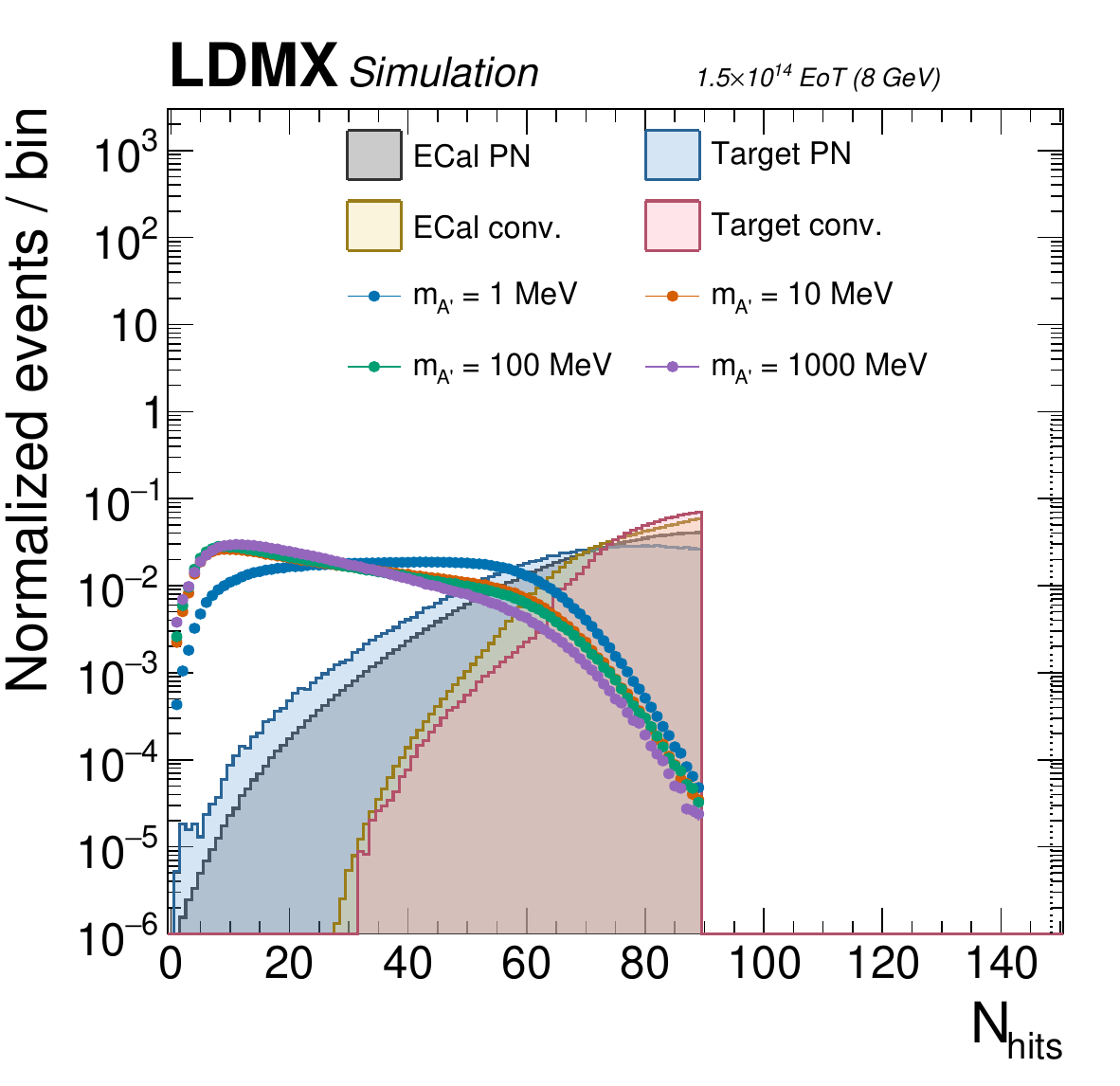}
    \caption{Left: BDT score after the trigger and preselection (selection at 0.9547). Right: number of
    ECal readout hits, used in the cut-and-count selection. Both use the v15 geometry.}
    \label{fig:bdt}
\end{figure}

The ECal backgrounds are removed with a boosted decision tree (BDT)~\cite{10.1145/2939672.2939785}
built from ECal shower and hit features (Fig.~\ref{fig:bdt}, left), with the selection at 0.9547.
Table~\ref{tab:cutflow_bdt} shows the cutflow: after the HCal veto, 1 background event remains in
total, while 72--79\% of the signal is retained. This is higher than the 50--68\% with two background events at
$10^{14}$\,EoT of the design report~\cite{LDMX:2025rwc} and Ref.~\cite{Vami:2025phep}, and no longer
requires the tracker veto or the MIP tracking inside the ECal.

\subsection{Alternative selections}
\label{sec:alternatives}

The alternative approaches to the BDT were also re-optimized; Table~\ref{tab:summary} compares the three selections.
The graph neural network ParticleNet~\cite{Qu:2019gqs}, applied after a loose preselection
(fewer than 90 hits, isolated ECal energy below 1100\,MeV) with a score above 0.5664, leaves 1.7
background events with 66--81\% signal efficiency, comparable to the BDT.

The cut-and-count analysis applies requirements on the BDT features ranked most important by
SHAP~\cite{lundberg2019consistentindividualizedfeatureattribution}, with thresholds re-optimized with
the Punzi figure of merit~\cite{Punzi:2003bu}. Signal
showers are shorter, more isolated, and have fewer hits (Fig.~\ref{fig:bdt}, right); the requirement
of fewer than 57 hits is the most powerful, reducing the ECal PN background from 188 to 8 events. With
fewer than two straight MIP tracks in the ECal, 3.1 background events remain with 62--69\% signal
efficiency.

\begin{table}[htbp]
\centering
\caption{Background yields at $1.5\times10^{14}$\,EoT and signal efficiencies after all requirements for the three selections.}
\label{tab:summary}
\small
\begin{tabular}{|l||c|c|c|c|c|} \hline
Selection & Bkg.& 1\,MeV & 10\,MeV & 100\,MeV & 1000\,MeV \\ \hline
BDT           & 1   & 71.7\% & 75.1\% & 78.7\% & 74.4\% \\ \hline
Cut and count & 3.1 & 64.2\% & 68.1\% & 69.3\% & 62.0\% \\ \hline
ParticleNet   & 1   & 81.2\% & 80.1\% & 79.9\% & 66.2\% \\ \hline
\end{tabular}
\end{table}

\section{Expected sensitivity and signal characterization}
\label{sec:reach}

\begin{figure}[htbp]
    \centering
    \includegraphics[height=4.8cm]{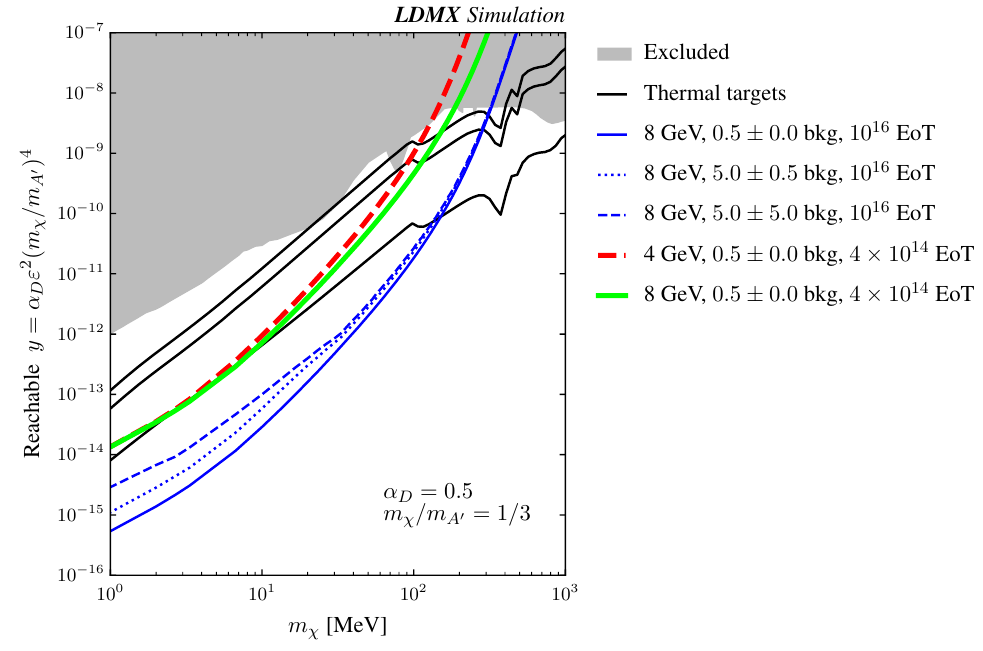}\hspace{0.6cm}%
    \includegraphics[height=4.8cm]{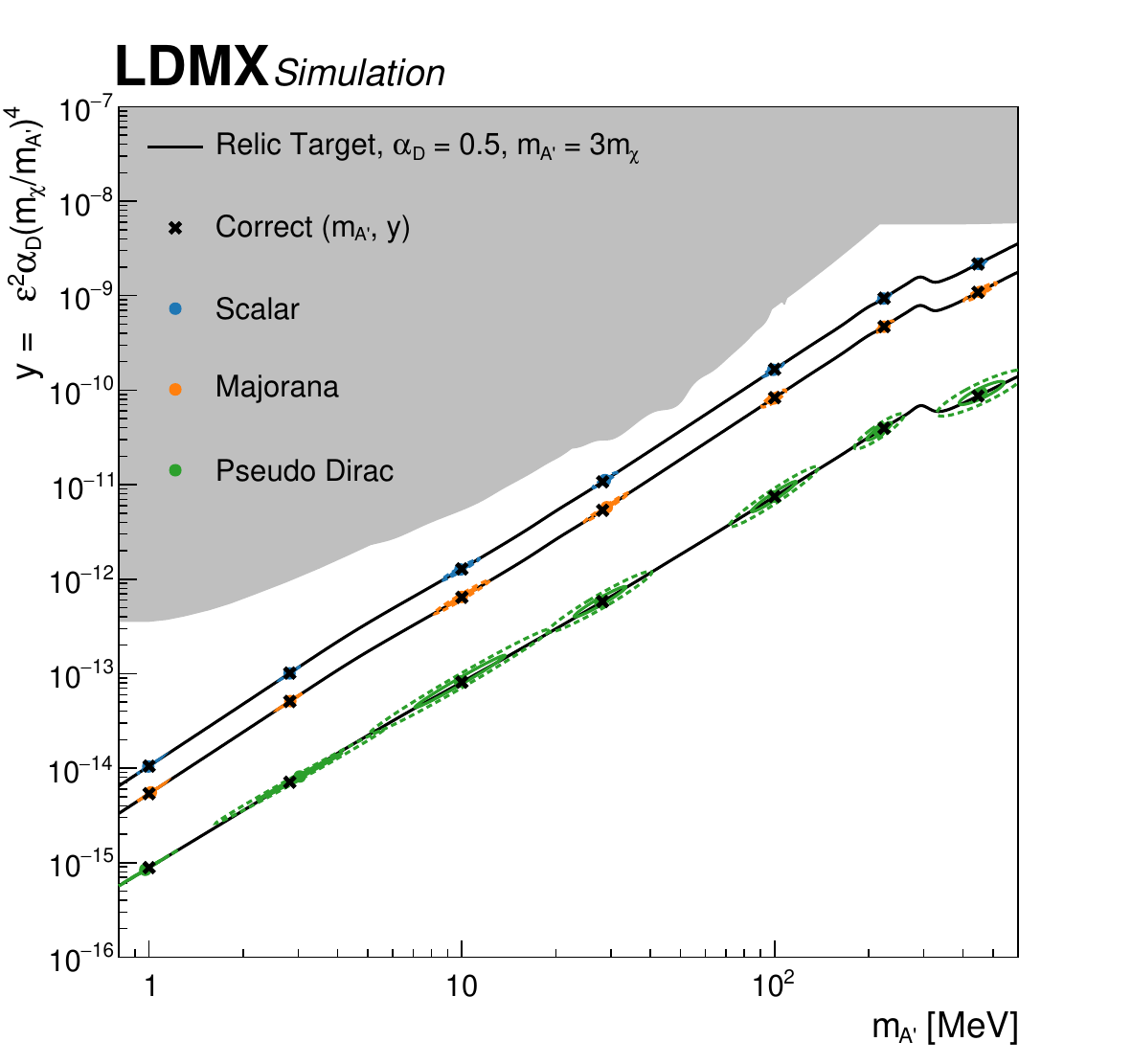}
    \caption{Left: expected limits for a range of remaining background events~\cite{LDMX:2023zbn}.
    Right: simulated (crosses) and reconstructed masses and couplings with $1\sigma$ (solid) and
    $2\sigma$ (dashed) ellipses for different models at $10^{16}$\,EoT~\cite{LDMX:2025rwc}.}
    \label{fig:reach}
\end{figure}

If no signal is observed, LDMX sets limits on the kinetic mixing parameter $\epsilon$, shown on the
left of Fig.~\ref{fig:reach} for a DM to $\mathrm{A}'$ mass ratio of 1/3 and a range of residual
background yields~\cite{LDMX:2023zbn}. A large part of the thermal targets (black lines) is covered
already with $4\times10^{14}$\,EoT, about one year of running.

If a signal is observed, the recoil electron $p_\mathrm{T}$, unused by the selections above,
characterizes it~\cite{Blinov:2020epi}. Modeling its spectrum with a Gaussian process regression
(including $p_\mathrm{T}$ smearing), a likelihood fit reconstructs
the $\mathrm{A}'$ mass and coupling with good precision~\cite{LDMX:2025rwc}, and with $10^{16}$\,EoT
distinguishes scalar, Majorana, and pseudo-Dirac DM (Fig.~\ref{fig:reach}, right).

\section{Beyond the missing-momentum search}
\label{sec:other}

LDMX is also a fully instrumented beam dump. A dedicated study~\cite{LDMX:2026llp} considers dark
photons and axion-like particles (ALPs) that decay visibly after traveling about a meter, producing a
displaced electromagnetic shower deep inside the HCal. Inverting the HCal veto (more than 4840\,MeV in the HCal) and
adding a second BDT on HCal features, the background is reduced from about 31\,000 events to fewer than one at $10^{14}$\,EoT, with a signal
efficiency of about 30\%. For ALPs coupling to electrons, LDMX has unique sensitivity between about 20
and 56\,MeV with $10^{16}$\,EoT.

Using the ECal as the target probes unexplored parameter space already in the first
data~\cite{LDMX:2025ixw}. LDMX can also extend the limits on millicharged fermions and other long-lived
models~\cite{Berlin_2019}, and its electronuclear measurements can reduce the electron-neutrino
modeling uncertainties relevant for neutrino physics.

\section{Vertical slice test and outlook}
\label{sec:status}

LCLS-II is being upgraded to 8\,GeV over the next one and a half years. Before the shutdown, LDMX
recorded some of the last 4\,GeV electrons in ESA with a ``vertical slice'': production-like modules of
every subsystem read out through the full trigger and DAQ chain. The test characterized the beam line into ESA and demonstrated combined track
reconstruction with the scintillating trigger bars and the silicon planes in beam data. The data are
being analyzed, and all-system integration tests with cosmic rays are in preparation.

\section{Summary}
\label{sec:summary}

LDMX will probe sub-GeV DM at SLAC through its missing-momentum signature, suppressing SM backgrounds
by more than 14 orders of magnitude. With the updated geometry, more realistic tracking, and three times more
simulated events, the BDT selection leaves 1 background events at $1.5\times10^{14}$\,EoT with
72--79\% signal efficiency. This
covers much of the thermal targets within a year. The same detector reaches new
ground for long-lived dark photons and ALPs, and a vertical slice has already recorded beam data: LDMX is
ready to be built as a DM discovery experiment.

{\footnotesize
\setlength{\bibsep}{2pt plus 0.5pt}

}

\end{document}